\documentstyle[12pt]{article}
\begin{document}
\renewcommand{\theequation}{\thesection.\arabic{equation}}
\vskip 2cm
\title{Second Order Perturbations of a Macroscopic String; Covariant Approach}
\author{\\
 A.L. Larsen${}^{|} {}^{1}$ and A. Nicolaidis${}^{||} {}^{2}$}
\maketitle
\noindent
$^{1}${\em Physics Department, University of Odense, Campusvej 55, 5230
Odense M,
Denmark}\\
$^{2}${\em Theoretical  Physics Department, University of Thessaloniki,
54006 Thessaloniki, Greece}
\noindent
\vskip 6cm
\noindent
$^{|}$Electronic address: all@fysik.sdu.dk\\
$^{||}$Electronic address: nicolaid@ccf.auth.gr
\newpage
\begin{abstract}
\baselineskip=1.5em
Using a world-sheet covariant formalism, we derive the equations of motion
for second order perturbations of a generic macroscopic string, thus
generalizing
previous results for first order perturbations. We give the explicit
results for
the first and second order perturbations of a contracting near-circular string;
these results are relevant for the understanding of the possible outcome when a
cosmic string contracts under its own tension, as discussed in a series of
papers
by Vilenkin and Garriga. In particular, second order perturbations are
necessaary
for a consistent computation of the energy.

We also quantize the perturbations and
derive the mass-formula up to second order in perturbations for an observer
using
world-sheet time $\tau $. The high frequency modes give the standard Minkowski
result while, interestingly enough, the Hamiltonian turns out to be
non-diagonal in oscillators for low-frequency modes. Using an alternative
definition of the vacuum, it is possible to diagonalize the Hamiltonian,
and the
standard string mass-spectrum appears for all frequencies. We finally
discuss how
our results are also relevant for the problems concerning string-spreading
near a
black hole horizon, as originally discussed by Susskind.
\hspace*{-6mm}\end{abstract}
\newpage
\section{Introduction}
\setcounter{equation}{0}
It is well known that the classical string equations of motion in flat
Minkowski
space can be solved exactly using conformal gauge (see for instance
\cite{green}).
Moreover, the gauge constraints arising in conformal gauge can be solved
exactly
by supplementing conformal gauge with light-cone gauge. However, for many
purposes,
especially in connection with macroscopic cosmic strings, the formalism of
conformal and light cone gauge, although mathematically tractable, is not
particularly useful. First of all, the world-sheet time $\tau$ is generally not
related  to the preferred coordinate time $t$ in a simple way. Secondly,
although
the longitudinal oscillations are expressed in terms of the transverse
oscillations, and therefore do not represent independent physical degrees of
freedom, the longitudinal oscillations are actually still present. Finally, for
macroscopic cosmic strings there is often a natural separation of the
degrees of
freedom into "slow modes" and "fast modes", and this separation is often more
transparent in alternative gauges. In a curved spacetime the situation is even
worse: The classical string equations of motion cannot generally be solved in
conformal gauge (nor in any other gauge)  and it is generally not even
possible to
supplement conformal gauge with light cone gauge, since it would be
inconsistent
with the equations of motion.

For macroscopic strings with small oscillations it is usually much more
convenient
to use a formalism where the small oscillations are considered as
perturbations.
Instead of conformal gauge one can make a more physical gauge-choice where,
from
the beginning, only transverse oscillations are present. Moreover, the
world-sheet
time $\tau$ of the unperturbed macroscopic string can be directly
identified with
the preferred coordinate time $t$. Another advantage is that spacetime and
world-sheet covariance can be maintained at all stages. The price to pay
for such
non-conformal gauge-choice is that the string equations of motion are
non-linear.
However, that is not really a problem in a perturbative scheme where the equations
of motion are to be solved order by order in the expansion around the zero
th
order unperturbed macroscopic string.

The world-sheet covariant perturbative approach was developed in
\cite{garriga} for membranes and strings in flat Minkowski space
and in de Sitter space (a non-covariant approach was previously developed in
\cite{san}). The results were generalized to arbitrary curved  backgrounds in
\cite{frolov,guven,carter} (See also \cite{cap} for some recent developments).
However, until now only first order perturbations around the zeroth order
macroscopic string have been considered. This is perfectly enough for many
purposes, but in certain cases it is necessary to consider also the second
order
perturbations. For instance, considering small perturbations around a
contracting
circular string, it is easy to see that there is no contribution to the total
conserved energy to first order; the first order contribution simply integrates
out. The first non-zero contribution (besides the zeroth order
contribution) to the
total energy is quadratic in the first order perturbations, but then also
second
order perturbations must be included for consistency, since they
contribute to the
same order.

The purpose of the present paper is first of all to generalize the results of
[2,4-6] for first order perturbations to second order perturbations. That
is, we
derive the equations of motion for the second order perturbations in
world-sheet
covariant form.  We then give the explicit results for
the first and second order perturbations of a contracting near-circular string;
these results are relevant for the understanding of the possible outcome when a
cosmic string contracts under its own tension \cite{garriga}. Moreover, as
already
mentioned, the second order perturbations are necessary also for a consistent
computation of the total conserved energy. After obtaining explicitly the
expression for the classical mass-energy,
we quantize the perturbations and
derive the quantum mass-formula up to second order in perturbations for an
observer
using world-sheet time $\tau$. The high frequency modes give rise to the
standard
Mikowski result while, interestingly enough, the Hamiltonian
turns out to be non-diagonal in oscillators for low-frequency modes. We
then show
that using an alternative definition of the vacuum, it is possible to
diagonalize
the Hamiltonian, and the standard string mass-spectrum appears for all
frequencies. We finally discuss how our results are also relevant for the
problems
concerning string-spreading near a black hole horizon, as originally
discussed by
Susskind
\cite{susk1}.

\section{General Formalism}
\setcounter{equation}{0}
Our starting point is the Nambu-Goto  action
\begin{equation}
S=\frac{1}{2\pi\alpha'}\int d\tau d\sigma\;\sqrt{-G}\; ;\;\;\;\;\;\;G
\equiv{\mbox{det}}(G_{AB})
\end{equation}
where $G_{AB}$ is the induced metric
on the string
world-sheet
\begin{equation}
G_{AB}\equiv \eta_{\mu\nu}x^\mu_{,A}x^\nu_{,B}
\end{equation}
Here $(A, B)=(0, 1)$ are  the world-sheet indices, while $(\mu, \nu)=(0, 1,
2, 3)$
are the
spacetime indices. We consider strings in 4-dimensional
Minkowski space using Cartesian coordinates and sign-conventions
$\eta_{\mu\nu}=
{\mbox{diag}}(-1,1,1,1)$.

The conjugate momentum $P^{A}_{\mu}$ is given by
\begin{equation}
P^{A}_{\mu}\equiv \frac{\delta{\cal L}}{\delta
x^{\mu}_{,A}}=\frac{1}{2\pi\alpha'}
\sqrt{-G}\;G^{AB}x_{\mu,B}
\end{equation}
The  equations of motion, corresponding to the action (2.1),
are then:
\begin{equation}
\partial_A P^{A}_{\mu}=0
\end{equation}
As is well-known, the tangential projection of eq.(2.4), $x^\mu_{,B}\partial_A
P^{A}_{\mu}=0$, is an identity, thus the equation of motion is equivalently
\begin{equation}
n^\mu_{i}\partial_A P^{A}_{\mu}=0
\end{equation}
where the two normal vectors $n^\mu_{i}$ $(i=1,2)$ are introduced by
\begin{equation}
\eta_{\mu\nu}n^\mu_{i}n^\nu_{j}=\delta_{ij}\; ,\;\;\;\;\;\;\;\;\eta_{\mu\nu}
n^\mu_{i}x^\nu_{,A}=0
\end{equation}
We shall be interested in physical situations where the solution to eq.(2.4)
(or, equivalently, eq.(2.5)) is naturally described as a macroscopic string
experiencing small perturbations. Up to second order perturbations, we
therefore
write $x^\mu$ in the following way
\begin{equation}
x^\mu=\bar{x}^\mu+\delta x^\mu_{(1)}+\delta x^\mu_{(2)}
\end{equation}
Here, and in the following, the bar represents unperturbed (zeroth order)
quantities.
 Moreover, since we are interested only
in physical
(transverse) perturbations, $\delta x^\mu$ can be expanded on the
normal vectors, $\delta x^\mu=n^\mu_{i}\Phi^{i}$. Thus, by expanding up to
second order
\begin{equation}
\delta x^\mu_{(1)}=\bar{n}^\mu_{i}\Phi^{i}_{(1)}
\end{equation}
\begin{equation}
\delta x^\mu_{(2)}=\bar{n}^\mu_{i}\Phi^{i}_{(2)}+\delta n^{\mu}_{i(1)}
\Phi^{i}_{(1)}
\end{equation}
In eq.(2.9) we need the first order perturbation of the normal vector,
$\delta n^{\mu}_{i(1)}$. It is easily obtained from eq.(2.6)
\begin{equation}
\delta n^{\mu}_{i(1)}=-(\bar{D}_{ij}\;^{A}\Phi^{j}_{(1)})\bar{x}^\mu_{,A}
\end{equation}
where the covariant derivative $\bar{D}_{ijA}$ is defined by
\begin{equation}
\bar{D}_{ijA}\equiv \delta_{ij}\bar{\nabla}_A+\bar{\mu}_{ijA}
\end{equation}
Here $\bar{\nabla}_{A}$ is the covariant derivative with respect to the
induced
metric $\bar{G}_{AB}$ on the unperturbed world-sheet, while
$\bar{\mu}_{ijA}$ is
the torsion (normal fundamental form) of the unperturbed world-sheet
\begin{equation}
\bar{\mu}_{ijA}\equiv \eta_{\mu\nu}\bar{n}^\mu_{i}  \bar{n}^\nu_{j,A}
\end{equation}
The equations of motion of the first and second order perturbations are
conveniently written in terms of geometric quantities such as the covariant
derivative $\bar{D}_{ijA}$ and the extrinsic curvature (second fundamental
form)
$\bar{\Omega}_{iAB}$ of the unperturbed world-sheet
\begin{equation}
\bar{\Omega}_{iAB}\equiv \eta_{\mu\nu}\bar{n}^\mu_{i}  \bar{x}^\nu_{,AB}
\end{equation}
This  makes manifest the world-sheet covariance as well as the SO(2)
invariance
under rotations of the normal vectors. Some useful formulas are given in the
appendix. Here we just list the results for the conjugate momentum
$P^{A}_{\mu}$.

To zeroth order, it is simply
\begin{equation}
\bar{P}^{A}_\mu=\frac{1}{2\pi\alpha'}\sqrt{-\bar{G}}\;\bar{G}^{AB}
\bar{x}_{\mu,B}
\end{equation}
The first order perturbation is also easily obtained
\begin{equation}
\delta P^{A}_{\mu(1)}=\frac{1}{2\pi\alpha'}\sqrt{-\bar{G}}\left( (\bar{D}_{ij}
\;^{A}\Phi^{j}_{(1)})\bar{n}^{i}_\mu+
(\bar{\Omega}_{i}\;^{AB}-\bar{G}^{AB}\bar{\Omega}_{iC}\;^{C})
\Phi^{i}_{(1)}\bar{x}_{\mu,B}\right)
\end{equation}
The second order perturbation is considerably more complicated. Using the
formulas
of Appendix A, it becomes
\begin{eqnarray}
\delta P^{A}_{\mu(2)}&=&\frac{1}{2\pi\alpha'}\sqrt{-\bar{G}}
\left( (\bar{D}_{ij}\;^{A}\Phi^{j}_{(2)})\bar{n}^{i}_\mu+
(\bar{\Omega}_{i}\;^{AB}-\bar{G}^{AB}
\bar{\Omega}_{iC}\;^{C})\Phi^{i}_{(2)}\bar{x}_{\mu,B}\right.
\nonumber\\
&+&(\bar{D}_{k}\;^{jB}\bar{D}_{ji}\;^{A}\Phi^{i}_{(1)})
\Phi^{k}_{(1)}\bar{x}_{\mu,B}-\bar{G}^{AB}
(\bar{D}_{k}\;^{jC}\bar{D}_{jiC}\Phi^{i}_{(1)})
\Phi^{k}_{(1)}\bar{x}_{\mu,B}\nonumber\\
&-&\frac{1}{2}\bar{G}^{AB}(\bar{D}_{kj}\;^{C}\Phi^{j}_{(1)})
(\bar{D}^{k}\;_{iC}
\Phi^{i}_{(1)})\bar{x}_{\mu,B}-(\bar{D}_{ik}\;^{A}
\Phi^{k}_{(1)})\bar{\Omega}_{jC}\;^{C}\Phi^{j}_{(1)}
\bar{n}_{\mu}^{i}
\nonumber\\
&+&2(\bar{D}_{ikB}\Phi^{k}_{(1)})\bar{\Omega}_{j}\;^{AB}
\Phi^{j}_{(1)}\bar{n}^{i}_{\mu}-(\bar{D}_{jkB}\Phi^{k}_{(1)})\bar{\Omega}_{i}
\;^{AB}
\Phi^{j}_{(1)}\bar{n}_{\mu}^{i}\nonumber\\
&+&\frac{1}{2}\bar{G}^{AB}(\bar{\Omega}_{jC}\;^{C}
\bar{\Omega}_{iD}\;^{D}-\bar{\Omega}_{jCD}
\bar{\Omega}_{i}\;^{CD})\Phi^{j}_{(1)}
\Phi^{i}_{(1)}\bar{x}_{\mu,B}\nonumber\\
&+&\left. (\bar{\Omega}_{j}\;^{AC}
\bar{\Omega}_{iC}\;^{B}-\bar{\Omega}_{jC}\;^{C}
\bar{\Omega}_{i}\;^{AB})\Phi^{j}_{(1)}
\Phi^{i}_{(1)}\bar{x}_{\mu,B}\right)
\end{eqnarray}
Now it is straightforward to obtain the equations of motion (2.4) order by
order
in the expansion. To zeroth order, it is simply
\begin{equation}
\bar{\Omega}_{iC}\;^{C}=0
\end{equation}
i.e., the well-known result of vanishing mean extrinsic curvature for a
minimal
surface.

To first order, the result is (using also the zeroth order equation of motion)
\begin{equation}
\left(\bar{D}_{jk}\;^{A}\bar{D}^{k}\;_{iA}+\bar{\Omega}_{jAB}
\bar{\Omega}_{i}\;^{AB}\right)\Phi^{i}_{(1)}=0
\end{equation}
as was already obtained independently in a number of papers
\cite{frolov,guven}.

The second order equation of motion, which to our knowledge has  not been
obtained
before in covariant form, becomes (using also the zeroth and first order
equations
of motion)
\begin{equation}
\left(\bar{D}_{jk}\;^{A}\bar{D}^{k}\;_{iA}+\bar{\Omega}_{jAB}
\bar{\Omega}_{i}\;^{AB}\right)\Phi^{i}_{(2)}=f_{j}
\end{equation}
where the source $f_{j}$ is given in terms of the first order perturbations
\begin{eqnarray}
f_{j}=&-&2(\bar{D}_{jiB}\bar{D}^{i}\;_{kA}\Phi^{k}_{(1)})
\bar{\Omega}_{l}\;^{AB}\Phi^{l}_{(1)}
-2(\bar{D}_{jkA}\Phi^{k}_{(1)})(\bar{D}^{i}\;_{lB}
\Phi^{l}_{(1)})\bar{\Omega}_{i}\;^{AB}\nonumber\\
&+&(\bar{D}^{i}\;_{kA}\Phi^{k}_{(1)})(\bar{D}_{ilB}
\Phi^{l}_{(1)})\bar{\Omega}_{j}\;^{AB}
\end{eqnarray}
It should be mentioned that the derivation of the sourceterm from
eqs.(2.4), (2.16) is a somewhat lengthy exercise in differential geometry.
To finally obtain the sourceterm in the relatively simple form, eq.(2.20),
we used among other things the completeness relation
\begin{equation}
\eta^{\mu\nu}=\bar{G}^{AB}\bar{x}^\mu_{,A}\bar{x}^\nu_{,B}+
\delta^{ij}\bar{n}^\mu_{i}\bar{n}^\nu_{j}
\end{equation}
the Weingarten equation
\begin{equation}
\bar{\nabla}_{A}\bar{\nabla}_{B}\bar{x}^\mu=
\bar{\Omega}^{i}\;_{AB}\bar{n}^\mu_{i}
\end{equation}
the Gauss-Codazzi equation
\begin{equation}
\bar{D}_{ijA}\bar{\Omega}^{jA}\;_{B}=\bar{D}_{ijB}
\bar{\Omega}^{jA}\;_{A}
\end{equation}
as well as the identity
\begin{eqnarray}
2\bar{\Omega}_{iA}\;^{B}\bar{\Omega}_{jB}\;^{C}
\bar{\Omega}_{kC}\;^{A}&=&\bar{\Omega}_{iA}\;^{B}
\bar{\Omega}_{jB}\;^{A}\bar{\Omega}_{kC}\;^{C}+
\bar{\Omega}_{jA}\;^{B}\bar{\Omega}_{kB}\;^{A}
\bar{\Omega}_{iC}\;^{C}\nonumber\\
&+&\bar{\Omega}_{kA}\;^{B}\bar{\Omega}_{iB}\;^{A}
\bar{\Omega}_{jC}\;^{C}-\bar{\Omega}_{iA}\;^{A}
\bar{\Omega}_{jB}\;^{B}\bar{\Omega}_{kC}\;^{C}
\end{eqnarray}
Notice in particular that the zeroth order equation of motion ensures
vanishing
of the left hand sides of eq.(2.23)-(2.24).
\section{Circular String}
\setcounter{equation}{0}
We now consider the case where the unperturbed string is a circular string
in the
$x-y$ plane. Defining $R_{0}$ to be the maximal radius, it is thus
parametrized by:
\begin{eqnarray}
\bar{t}&=&R_{0}\tau\nonumber\\
 \bar{x}&=&R_{0}\cos\tau\cos\sigma\\
 \bar{y}&=&R_{0}\cos\tau\sin\sigma\nonumber\\
\bar{z}&=&0\nonumber
\end{eqnarray}
such that
\begin{equation}
\bar{G}_{AB}={\mbox{diag}}(-R_{0}^2\cos^2\tau, \;R_{0}^2\cos^2\tau)
\end{equation}
The unperturbed normal vectors are given by
\begin{equation}
\bar{n}^\mu_{1}=\left( \begin{array}{c}
\sin\tau / \cos\tau \\
-\cos\sigma / \cos\tau \\ -\sin\sigma / \cos \tau \\
0 \end{array}  \right) \; ,\;\;\;\;\
\bar{n}^\mu_{2}=\left( \begin{array}{c}
0 \\
0 \\
0\\
1 \end{array}  \right)
\end{equation}
The only non-vanishing components of the unperturbed extrinsic curvature
$\bar{\Omega}_{iAB}$ are
\begin{equation}
\bar{\Omega}_{1\tau\tau}=\bar{\Omega}_{1\sigma\sigma}=R_{0}
\end{equation}
while all components of the unperturbed torsion
$\bar{\mu}_{ijA}$ vanish.

From eqs.(3.2)-(3.4) follows that the zeroth order equation of motion,
eq.(2.17),
is trivially fulfilled.

The first order equation of motion, eq.(2.18), reduces to
\begin{equation}
\left( -\partial^2_\tau+\partial^2_\sigma+\frac{2}{\cos^2\tau}\right)
\Phi^{1}_{(1)}=0
\end{equation}
\begin{equation}
\left( -\partial^2_\tau+\partial^2_\sigma\right) \Phi^{2}_{(1)}=0
\end{equation}
These equations are easily solved using Fourier expansions
\begin{equation}
\Phi^{i}_{(1)}(\tau,\sigma)=\sqrt{\frac{\alpha'}{2}}
\sum_{n=-\infty}^{\infty}C^{i}_{n}(\tau)e^{-in\sigma}\;\; ;\;\;\;\; C^{i}_{-n}=
C^{i*}_{n}
\end{equation}
The Fourier coefficients $C^{1}_{n}$ are given by
\begin{eqnarray}
C^{1}_{0}&=&\sqrt{2}\left( a^{1}_{0}(1+\tau\tan\tau)+b^{1}_{0}\tan\tau\right)
\nonumber\\
C^{1}_{1}&=&a^{1}_{1}\left( \frac{1}{\cos\tau}-\frac{i}{2}\left( \frac{\tau}
{\cos\tau}+\sin\tau\right)\right)+
b^{1*}_{1}\left( \frac{1}{\cos\tau}+\frac{i}{2}\left( \frac{\tau}{\cos\tau}+
\sin\tau\right)\right)\nonumber\\
C^{1}_{n}&=&a^{1}_{n}\frac{n+i\tan\tau}{\sqrt{n(n^2-1)}}
\;e^{-in\tau}+b^{1*}_{n}\frac{n-i\tan\tau}{\sqrt{n(n^2-1)}}
\;e^{in\tau}\;\; ;\;\;\;\;n\geq 2
\end{eqnarray}
The Fourier coefficients $C^{2}_{n}$ are given by
\begin{eqnarray}
C^{2}_{0}&=&\sqrt{2}\left( a^{2}_{0}+b^{2}_{0}\tau\right) \nonumber\\
C^{2}_{1}&=&a^{2}_{1}e^{-i\tau}+
b^{2*}_{1}e^{i\tau}\nonumber\\
C^{2}_{n}&=&\frac{1}{\sqrt{n}}a^{2}_{n}e^{-in\tau}+
\frac{1}{\sqrt{n}}b^{2*}_{n}e^{in\tau}\;\; ;\;\;\;\;n\geq 2
\end{eqnarray}
Recall that the negative $n$ modes are defined in terms of the above ones by
$C^{i}_{-n}=C^{i*}_{n}$. Notice also that the $n=0,\pm 1$ modes are treated
separately. This will be explained later; see also \cite{garriga}.
Moreover, the
normalizations and precise definitions of the modes $a^{i}_{n}, b^{j}_{m}$ in
eqs.(3.8)-(3.9) are motivated as follows: The first order equations of motion,
eqs.(3.5)-(3.6), correspond to the effective action
\begin{equation}
{\cal S}=\frac{1}{4\pi\alpha'}\int d\tau d\sigma\left\{ \Phi^{1}_{(1)}
\left(  -\partial^2_\tau+\partial^2_\sigma+\frac{2}{\cos^2\tau}\right)
\Phi^{1}_{(1)}+\Phi^{2}_{(1)}\left(  -\partial^2_\tau+\partial^2_\sigma\right)
\Phi^{2}_{(1)}\right\}
\end{equation}
The conjugate momenta are defined by $\Pi^{i}_{(1)}
\equiv \delta {\cal S}/ \delta \dot{\Phi}^{i}_{(1)}=\dot{\Phi}^{i}_{(1)}/2\pi
\alpha'$.
They correspond to normal projections of the conjugate momenta $\delta
P^\tau_{\mu(1)}$ introduced in eq.(2.15). Then the canonical Poisson
brackets are
\begin{equation}
\left\{ \Pi^{i}_{(1)}(\tau,\sigma),\;\Phi^{j}_{(1)}(\tau,\sigma')\right\}=
-\delta^{ij}\delta (\sigma-\sigma')
\end{equation}
and it is straightforward to show that the modes $a^{i}_{n}, b^{i}_{n}$ obey
\begin{eqnarray}
&\left\{ b^{i}_{0}, a^{j}_{0}\right\}&=-\delta^{ij}\nonumber\\
&\left\{ a^{i}_{n},a^{j*}_{m}\right\}&=\left\{ b^{i}_{n},b^{j*}_{m}\right\}=
-i\delta^{ij}\delta_{nm}\; ; \;\;\; n,m
\geq 1
\end{eqnarray}
i.e. $a^{i}_{0}, b^{j}_{0}$ are conventionally normalized center of mass
coordinate
and momenta, while $a^{i}_{n}, b^{j}_{m}$ ($n,m\geq 1$) are conventionally
normalized harmonic oscillator modes.

We now come to the second order equation of motion, eq.(2.19). In the
present case, it reduces to
\begin{equation}
\left( -\partial^2_\tau+\partial^2_\sigma+\frac{2}{\cos^2\tau}\right)
\Phi^{1}_{(2)}=R^2_{0}\cos^2\tau f^{1}
\end{equation}
\begin{equation}
\left( -\partial^2_\tau+\partial^2_\sigma\right) \Phi^{2}_{(2)}=
R^2_{0}\cos^2\tau f^{2}
\end{equation}
where the sourceterms $f^{i}$, eq.(2.20), are given by
\begin{eqnarray}
R^2_{0}\cos^2\tau f^{1}&=&\frac{-2}{R_{0}\cos^2\tau}\left(
\Phi^{1}_{(1)}\ddot{\Phi}^{1}_{(1)}+\Phi^{1}_{(1)}\Phi''^{1}_{(1)}
+\frac{2\sin\tau}{\cos\tau}\Phi^{1}_{(1)}\dot{\Phi}^{1}_{(1)}\right)
\nonumber\\
&-&\frac{1}{R_{0}\cos^2\tau}\left(
\dot{\Phi}^{1}_{(1)}\dot{\Phi}^{1}_{(1)}-
\dot{\Phi}^{2}_{(1)}\dot{\Phi}^{2}_{(1)}+
\Phi'^{1}_{(1)}\Phi'^{1}_{(1)}-\Phi'^{2}_{(1)}\Phi'^{2}_{(1)}\right)
\end{eqnarray}
\begin{eqnarray}
R^2_{0}\cos^2\tau f^{2}&=&\frac{-2}{R_{0}\cos^2\tau}\left(
\Phi^{1}_{(1)}\ddot{\Phi}^{2}_{(1)}+\Phi^{1}_{(1)}\Phi''^{2}_{(1)}
+\frac{2\sin\tau}{\cos\tau}\Phi^{1}_{(1)}\dot{\Phi}^{2}_{(1)}\right)
\nonumber\\
&-&\frac{2}{R_{0}\cos^2\tau}\left(
\dot{\Phi}^{1}_{(1)}\dot{\Phi}^{2}_{(1)}+
\Phi'^{1}_{(1)}\Phi'^{2}_{(1)}\right)
\end{eqnarray}
Eqs.(3.13)-(3.14) are solved using Fourier expansions
\begin{equation}
\Phi^{i}_{(2)}(\tau,\sigma)=\sqrt{\frac{\alpha'}{2}}
\sum_{n=-\infty}^{\infty}D^{i}_{n}(\tau)e^{-in\sigma}\;\; ;\;\;\;\; D^{i}_{-n}=
D^{i*}_{n}
\end{equation}
\begin{equation}
-R^2_{0}\cos^2\tau f^{i}(\tau,\sigma)=\sqrt{\frac{\alpha'}{2}}
\sum_{n=-\infty}^{\infty}
f^{i}_{n}(\tau)e^{-in\sigma}\;\; ;\;\;\;\; f^{i}_{-n}=f^{i*}_{n}
\end{equation}
(the factor $-R^2_{0}\cos^2\tau $ is included in $f^{i}$ for convenience).
Then eqs.(3.13)-(3.14) become
\begin{equation}
\ddot{D}^{1}_{n}+\left( n^2-\frac{2}{\cos^2\tau}\right) D_{n}^{1}=f_{n}^{1}
\end{equation}
\begin{equation}
\ddot{D}^{2}_{n}+n^2D_{n}^{2}=f_{n}^{2}
\end{equation}
where the sourceterms $f^{i}_{n}$ are given by
\begin{eqnarray}
f^{1}_{n}&=&\frac{\sqrt{2\alpha'}}{R_{0}\cos^2\tau}
\sum_{m=-\infty}^{\infty}\left\{C^{1}_{m}\left(\ddot{C}^{1}_{n-m}-
(n-m)(n-\frac{m}{2})C^{1}_{n-m}+\frac{2\sin\tau}{\cos\tau}
\dot{C}^{1}_{n-m}\right)
\right.
\nonumber\\
&+&\left. \frac{1}{2}\left(\dot{C}^{1}_{m}\dot{C}^{1}_{n-m}-
\dot{C}^{2}_{m}\dot{C}^{2}_{n-m}\right)+
\frac{m(n-m)}{2}{C}^{2}_{m}{C}^{2}_{n-m}\right\}
\end{eqnarray}
\begin{equation}
f^{2}_{n}=\frac{\sqrt{2\alpha'}}{R_{0}\cos^2\tau}
\sum_{m=-\infty}^{\infty}\left\{C^{1}_{m}\left(\ddot{C}^{2}_{n-m}-
n(n-m)C^{2}_{n-m}+\frac{2\sin\tau}{\cos\tau}
\dot{C}^{2}_{n-m}\right)
+\dot{C}^{1}_{m}\dot{C}^{2}_{n-m}\right\}
\end{equation}
By solving eq.(3.19), we find that the Fourier coefficients $D^{1}_{n}$ are
given by
\begin{eqnarray}
D^{1}_{0}&=&\tan\tau\int_0^\tau
(1+\tau'\tan\tau')f^{1}_{0}d\tau'-(1+\tau\tan\tau)
\int_0^\tau\tan\tau'f^{1}_{0}d\tau'\nonumber\\
&+&\sqrt{2}\left( \alpha^{1}_{0}(1+\tau\tan\tau)+\beta^{1}_{0}\tan\tau\right)
\nonumber\\
D^{1}_{1}&=&\frac{1}{2i}\left( \frac{1}{\cos\tau}+\frac{i}{2}\left( \frac{\tau}
{\cos\tau}+\sin\tau\right)\right)\int_0^\tau
\left( \frac{1}{\cos\tau'}-\frac{i}{2}\left( \frac{\tau'}{\cos\tau'}+\sin
\tau'\right)\right)f^{1}_{1}d\tau'
\nonumber\\
&-&\frac{1}{2i}\left( \frac{1}{\cos\tau}-\frac{i}{2}\left(
\frac{\tau}{\cos\tau}+
\sin\tau\right)\right)\int_0^\tau
\left( \frac{1}{\cos\tau'}+\frac{i}{2}\left( \frac{\tau'}{\cos\tau'}+\sin\tau'
\right)\right)f^{1}_{1}d\tau'
\nonumber\\
&+&\alpha^{1}_{1}\left( \frac{1}{\cos\tau}-\frac{i}{2}\left(
\frac{\tau}{\cos\tau}+
\sin\tau\right)\right)+
\beta^{1*}_{1}\left( \frac{1}{\cos\tau}+\frac{i}{2}\left(
\frac{\tau}{\cos\tau}+
\sin\tau\right)\right)\nonumber\\
D^{1}_{n}&=&\frac{n-i\tan\tau}{2in(n^2-1)}e^{in\tau}
\int_0^\tau(n+i\tan\tau')e^{-in\tau'}f^{1}_{n}d\tau'\nonumber\\
&-&\frac{n+i\tan\tau}{2in(n^2-1)}e^{-in\tau}\int_0^\tau(n-i\tan\tau')
e^{in\tau'}f^{1}_{n}d\tau'\nonumber\\
&+&\alpha^{1}_{n}\frac{n+i\tan\tau}{\sqrt{n(n^2-1)}}
\;e^{-in\tau}+\beta^{1*}_{n}\frac{n-i\tan\tau}{\sqrt{n(n^2-1)}}
\;e^{in\tau}\;\; ;\;\;\;\;n\geq 2
\end{eqnarray}
By solving eq.(3.20), we find that the Fourier coefficients $D^{2}_{n}$ are
given
by
\begin{eqnarray}
D^{2}_{0}&=&\tau\int_0^\tau f^{2}_{0}d\tau'-\int_0^\tau \tau'f^{2}_{0}d\tau'
+\sqrt{2}\left( \alpha^{2}_{0}+\beta^{2}_{0}\tau\right) \nonumber\\
D^{2}_{1}&=&\frac{1}{2i}e^{i\tau}\int_0^\tau
e^{-i\tau'}f^{2}_{1}d\tau'-\frac{1}{2i}e^{-i\tau}\int_0^\tau
e^{i\tau'}f^{2}_{1}d\tau'+\alpha^{2}_{1}e^{-i\tau}+\beta^{2*}_{1}
e^{i\tau}
\nonumber\\
D^{2}_{n}&=&\frac{1}{2in}e^{in\tau}\int_0^\tau
e^{-in\tau'}f^{2}_{n}d\tau'-\frac{1}{2in}e^{-in\tau}\int_0^\tau
e^{in\tau'}f^{2}_{n}d\tau'\nonumber\\
&+&\frac{1}{\sqrt{n}}\alpha^{2}_{n}\;e^{-in\tau}+
\frac{1}{\sqrt{n}}\beta^{2*}_{n}
\;e^{in\tau}\;\; ;\;\;\;\;n\geq 2
\end{eqnarray}
Again, we recall that the negative $n$ modes are defined in terms of the
above ones
by $D^{i}_{-n}=D^{i*}_{n}$.

This completes the exact explicit solution of the first and second order
perturbations around the circular string.
\section{ Mass-Spectrum}
\setcounter{equation}{0}
In this section we shall consider some physical quantities for the string,
including energy $\cal{E}$, momentum $\vec{P}$ and angular momentum
$T^\mu\;_\nu$.
They are all given in terms of the conjugate momentum introduced in eq.(2.3)
\begin{equation}
{\cal{E}}\equiv \int_0^{2\pi}d\sigma\;P^\tau_{t}
\end{equation}
\begin{equation}
\vec{P}\equiv \int_0^{2\pi}d\sigma\;\vec{P}^\tau
\end{equation}
\begin{equation}
T^\mu\;_\nu\equiv \int_0^{2\pi}d\sigma\;(x^\mu P^\tau_\nu-x_\nu P^{\tau\mu})
\end{equation}
and they are all conserved
\begin{equation}
\dot{\cal{E}}=0,\;\;\;\dot{\vec{P}}=0,\;\;\;\dot{T}^\mu\;_\nu=0
\end{equation}
as follows from eq.(2.4).

We shall also obtain the mass spectrum from the classical mass formula $M^2=
{\cal{E}}^2-\vec{P}^2$. Up to second order perturbations, it reads
\begin{eqnarray}
M^2&=&(\bar{{\cal{E}}}+\delta {\cal{E}}_{(1)}+\delta {\cal{E}}_{(2)}+...)^2-
(\vec{\bar{P}}+\delta \vec{P}_{(1)}+
\delta \vec{P}_{(2)}+...)^2\nonumber\\
&=&(\bar{{\cal{E}}}^2-\vec{\bar{P}}^2)+2(\bar{{\cal{E}}}
\delta {\cal{E}}_{(1)}-\vec{\bar{P}}\delta \vec{P}_{(1)})\nonumber\\
&+&(2\bar{{\cal{E}}}
\delta {\cal{E}}_{(2)}+\delta {\cal{E}}_{(1)}\delta {\cal{E}}_{(1)}-2
\vec{\bar{P}}\delta \vec{P}_{(2)}-\delta \vec{P}_{(1)}\delta \vec{P}_{(1)})+...
\end{eqnarray}
The physical quantities, of course, will depend on the particular choice of
initial
 conditions. The initial conditions are fixed in terms of the maximal
radius $R_0$
(zeroth order), the modes $a^{i}_{n}, b^{j}_{m}$ (first order) and the modes
$\alpha^{i}_{n}, \beta^{j}_{m}$ (second order).

To zeroth order, we get from eq.(2.3) and eqs.(3.1)-(3.2)
\begin{equation}
\bar{P}^\tau_\mu=\frac{-1}{2\pi\alpha'}\dot{\bar{x}}_\mu
\end{equation}
such that
\begin{equation}
\bar{{\cal{E}}}=R_0/\alpha',\;\;\;
\vec{\bar{P}}=0,\;\;\;\bar{T}^\mu\;_\nu=0
\end{equation}
i.e., only a zeroth order contribution to the energy.

To first order we get
\begin{equation}
\delta P^\tau_{\mu(1)}=\frac{-1}{2\pi\alpha'}\left(
\delta_{ij}\dot{\Phi}^{i}_{(1)}
\bar{n}^{j}_{\mu}-\frac{1}{R_0\cos^2\tau}\Phi^{1}_{(1)}\dot{\bar{x}}_\mu\right)
\end{equation}
from which we can easily obtain the first order contributions to ${\cal{E}}$,
$\vec{P}$ and $T^\mu\;_\nu$. However, following the discussion of
\cite{garriga},
we
take as initial conditions for the first order perturbations
\begin{equation}
a^{i}_0=b^{i}_0=a^{i}_{\pm 1}=b^{i}_{\pm 1}=0;\;\;\;i=1,2
\end{equation}
This is motivated by the fact that the $n=0,\pm 1$ modes correspond to rigid
spacetime translations and rotations of the circular string \cite{garriga}.
That is
to say, they do not really correspond to actual perturbations but merely to
redefinitions of the initial circular string. Then, using eq.(4.9), it is
easy to
show that
\begin{equation}
\delta {\cal{E}}_{(1)}=0,\;\;\;
\delta\vec{{P}}_{(1)}=0,\;\;\;\delta{T}^\mu\;_{\nu(1)}=0
\end{equation}
i.e., no contributions at all to the physical quantities from the first order
perturbations.

The second order contributions to the physical quantities can be
straightforwardly
computed, but here we shall give  only the result for the mass squared. As
follows
from eq.(4.5) and taking into account eqs.(4.7), (4.10), it is necessary to
compute
only $\delta {\cal{E}}_{(2)}$. Using eq.(2.16), we find that
\begin{eqnarray}
\delta P^\tau_{t(2)}&=&\frac{1}{2\pi\alpha'}\left(\frac{\sin\tau}
{\cos\tau}\dot{\Phi}^{1}_{(2)}-\frac{1}{\cos^2\tau}\Phi^{1}_{(2)}-
\frac{2\sin\tau}{R_0\cos^3\tau}\Phi^1_{(1)}\dot{\Phi}^1_{(1)}\right.
\nonumber\\
&+&\frac{2}{R_0\cos^4\tau}\Phi^1_{(1)}\Phi^1_{(1)}-\frac{1}{R_0
\cos^2\tau}(\Phi^1_{(1)}\ddot{\Phi}^1_{(1)}+
\Phi^2_{(1)}\ddot{\Phi}^2_{(1)})\nonumber\\
&+&\left.\frac{1}{2R_0\cos^2\tau}(\dot{\Phi}^1_{(1)}
\dot{\Phi}^1_{(1)}+
\dot{\Phi}^2_{(1)}\dot{\Phi}^2_{(1)}-\Phi'^{1}_{(1)}\Phi'^{1}_{(1)}-
\Phi'^{2}_{(1)}\Phi'^{2}_{(1)})\right)
\end{eqnarray}
such that the second order contribution to the energy, after some
straightforward
but tedious algebra,  becomes
\begin{eqnarray}
\delta {\cal{E}}_{(2)}&=&\frac{1}{\alpha'}\left(-\sqrt{\alpha'}\;
\alpha^{1}_{0}+\frac{\alpha'}{R_0}\sum_{n=2}^{\infty}n(a^{2*}_n
a^{2}_n+b^{2*}_n
b^{2}_n)\right.\\
&+&\left.\frac{\alpha'}{R_0}\sum_{n=2}^{\infty}\left(
\frac{2n^4-2n^2+1}{2n(n^2-1)}(a^{1*}_n a^{1}_n+b^{1*}_n b^{1}_n)+
\frac{2n^2-1}{2n(n^2-1)}(a^1_n b^{1}_n+a^{1*}_n
b^{1*}_n)\right)\right)\nonumber
\end{eqnarray}
The second order zero mode $\alpha^{1}_0$ can be set to zero for the same
reason as in eq.(4.9). Then we get the following result for the mass
formula up to second order perturbations
\begin{eqnarray}
M^2\alpha'&=&\frac{R^2_0}{\alpha'}+2\sum_{n=2}^{\infty}n
(a^{2*}_n a^{2}_n+b^{2*}_n b^{2}_n)\\
&+&2\sum_{n=2}^{\infty}\left(
\frac{2n^4-2n^2+1}{2n(n^2-1)}(a^{1*}_n a^{1}_n+b^{1*}_n b^{1}_n)+
\frac{2n^2-1}{2n(n^2-1)}(a^1_n b^{1}_n+a^{1*}_n b^{1*}_n)\right)\nonumber
\end{eqnarray}
This is the mass formula for an observer using stringtime $\tau$, i.e. for an
observer for which $a_n^i$ and $b_n^i$ are the positive frequency modes.
For large
$n$ (high frequency), the result reduces to the standard Minkowski result.
This
was to be expected since the high frequency modes do not "feel" that they are
living on a contracting  macroscopic circular string. For small $n$ (low
frequency), the result however differs significantly from the standard
Minkowski
result. Notice in particular that there are off-diagonal terms in the mass
formula.

It is convenient to introduce an alternative set of modes which
diagonalizes the
mass formula. We take $\tilde{a}_n^2 =a_n^2$ and $\tilde{b}_n^2 = b_n^2$ but
\begin{eqnarray}
\tilde{a}_n^1 &=& \frac{1}{2} \sqrt{\frac{n^2}{2(n^2-1)}} \left[
\frac{2n^2-1}{n^2}(a_n^1+b_n^1)+\frac{1}{n^2}(a_n^{1*}+b_n^{1*}) \right]\\
\tilde{b}_n^1 &=& \frac{-i}{2} \sqrt{\frac{n^2}{2(n^2-1)}} \left[
\frac{2n^2-1}{n^2}(a_n^1-b_n^1)-\frac{1}{n^2}(a_n^{1*}-b_n^{1*}) \right]
\end{eqnarray}
Then eq.(4.13) reduces to
\begin{equation}
M^2 \alpha ^\prime =\frac{R_0^2}{\alpha ^\prime} + 2\sum_{n=2}^{\infty}
n(\tilde{a}_n^{i*}
\tilde{a}_n^i+\tilde{b}_n^{i*} \tilde{b}_n^i)
\end{equation}
which is the standard Minkowski result, except for the "zero-point energy"
$\frac{R_0^2}{\alpha ^\prime}$ which is due to the unperturbed circular string.

The modes (4.14)-(4.15) fulfil the same algebra as in (3.12), but it is
important to
stress that they are \textbf{not} positive frequency modes for an observer
using
string time $\tau$. At the quantum level, it means that the vacuum defined
in terms
of the annihilation operators $\tilde{a}_n^i$,$\tilde{b}_n^i$
\begin{equation}
\tilde{a}_n^i|O>=\tilde{b}_n^i|O>=0
\end{equation}
does not coincide with the vacuum defined in terms of $a_n^i$,$b_n^i$. But
this is
of course a well-known problem when dealing with quantization of fields in
a curved
spacetime (see for instance \cite{davis}); in our case the fields are
represented by
the pertubations $\Phi ^i$ while the curved spacetime is represented by the
unperturbed contracting circular string.

\section{Discussion}
\setcounter{equation}{0}
 We considered the motion of a macroscopic string in flat Minkowski
spacetime. A string system is endowed with conformal symmetry on the
world-sheet and the string equations of motion are supplied with corresponding
constraints. The symmetry and the constraints underline the fact that only the
transverse string motion is physical. Therefore the actual physical degrees of
freedom of a string are $D-2$, rather than $D$, where $D$ is the dimension of
spacetime. There are two alternatives open: Either  work with all $D$
fields and
check at every stage of the calculation that the constraints are satisfied;
or work
directly with the transverse $D-2$ fields.

           Our case study, a macroscopic string contracting under the influence
of its own tension and experiencing small oscillations is most suited for the
second approach, as explained in the Introduction. We adopted the perturbative
scheme of Garriga and Vilenkin
\cite{garriga}. An exact special solution is introduced as the zeroth order
solution. The perturbation of the zeroth order solution should include only
tranverse oscillations and therefore lives in the subspace normal to the
unperturbed
string world-sheet. We improved the perturbative expansion by including second
order terms. Since at each  successive order the string world-sheet is
redefined, we have to redefine also the normal vectors to the world-sheet. Our
expressions for the second order perturbation are compact and are expressed
covariantly in terms of geometric quantities.

  Our formalism was exemplified by studying an almost circular string
in Minkowski spacetime. To zeroth order we have a circular string in the $x-y$
plane. We analyzed  to first and second order the transverse oscillations:
radial oscillations and oscillations in the $z$-direction. The $z$-oscillations
appear quite standard (the usual oscillations in flat spacetime). On the other
hand, the radial oscillations present novel features. The frequency spectrum of
the radial oscillations differs from the standard one for small and moderate
frequencies. Furthermore, a non-diagonal mixing  appears in the  mass
spectrum. This reflects  that the scalar  field (which represents the
perturbation) in the radial direction  is "feeling"  the underlying
two-dimensional contracting geometry. This is the picture emerging for an
observer
using world-sheet time $\tau$. However, by using an alternative definition
of the
vacuum, it is possible to diagonalize  the Hamiltonian, and the standard string
mass-spectrum appears for all frequencies.

 It is highly interesting to apply our formalism to strings moving
in curved spacetime. Along these lines we studied already \cite{nic} an
oscillating
circular string in Schwarzchild background to zeroth and first order. Our
motivation was to establish a framework within which to study in a precise
manner the string behavior near the black hole horizon, issues first raised by
Susskind \cite{susk1} (see also \cite{mezh}). We calculated both the radial
and angular spreading of the string, as the string approaches the black hole
horizon. We found that the radial spreading is suppressed by the Lorentz
contraction and the string appears (to an asymptotic observer) as wrapping
around
the event horizon. We plan to calculate and include the second order terms
and thus
analyze how the string oscillators spread over the event horizon. Notice
that the second order perturbations are necessary for a consistent
discussion of the energy. Hopefully we might understand the entropy of the
black
hole in terms of string degrees of freedom.

\vskip 1.5 cm
{\bf Acknowledgements}:\\
This work was supported by the Greek-Danish exchange program (Greek
Ministry of Education - Danish Rektorkollegiet). A.L. Larsen would like
to thank  for the warm hospitality
extended to him during his stay in Thessaloniki.
\newpage
\section{Appendix}
\setcounter{equation}{0}
The first and second order perturbations of $x^\mu$ are
\begin{eqnarray}
\delta x^\mu_{(1)}&=&\bar{n}^\mu_{i}\Phi^{i}_{(1)}\\
\delta x^\mu_{(2)}&=&\bar{n}^\mu_{i}\Phi^{i}_{(2)}-
(\bar{D}_{ij}\;^{A}\Phi^{j}_{(1)})\Phi^{i}_{(1)}\bar{x}^\mu_{,A}
\end{eqnarray}
It follows that
\begin{eqnarray}
(\delta x^\mu_{(1)})_{,B}&=&(\bar{D}^{i}\;_{jB}\Phi^{j}_{(1)})
\bar{n}^\mu_{i}-
\bar{\Omega}_{jB}\;^{C}\Phi^{j}_{(1)}\bar{x}^\mu_{,C}\\
(\delta x^\mu_{(2)})_{,B}&=&(\bar{D}^{i}\;_{jB}\Phi^{j}_{(2)})
\bar{n}^\mu_{i}-
\bar{\Omega}_{jB}\;^{C}\Phi^{j}_{(2)}\bar{x}^\mu_{,C}-
(\bar{D}_{ji}\;^{A}\Phi^{i}_{(1)})\bar{\Omega}^{k}\;_{AB}
\Phi^{j}_{(1)}\bar{n}^\mu_{k}\nonumber\\
&-&(\bar{D}_{ji}\;^{A}\Phi^{i}_{(1)})(\bar{D}^{j}\;_{kB}
\Phi^{k}_{(1)})\bar{x}^\mu_{,A}-(\bar{D}_{k}\;^{j}\;_{B}
\bar{D}_{ji}\;^{A}\Phi^{i}_{(1)})\Phi^{k}_{(1)}\bar{x}^\mu_{,A}
\end{eqnarray}
As for the first and second order perturbations of the (inverse)
induced metric on the world-sheet, we get
\begin{eqnarray}
\delta G^{AB}_{(1)}&=&2\bar{\Omega}_j\;^{AB}\Phi^{j}_{(1)}\\
\delta G^{AB}_{(2)}&=&2\bar{\Omega}_j\;^{AB}\Phi^{j}_{(2)}+
(\bar{D}_{ji}\;^{A}\Phi^{i}_{(1)})(\bar{D}^{j}\;_{k}\;^{B}
\Phi^{k}_{(1)})+3\bar{\Omega}_{j}\;^{A}\;_{C}
\bar{\Omega}_{i}\;^{BC}\Phi^{j}_{(1)}\Phi^{i}_{(1)}\nonumber\\
&+&(\bar{D}_{jk}\;^{A}\bar{D}^{k}\;_{i}\;^{B}\Phi^{i}_{(1)})
\Phi^{j}_{(1)}+(\bar{D}_{jk}\;^{B}\bar{D}^{k}\;_{i}\;^{A}
\Phi^{i}_{(1)})\Phi^{j}_{(1)}
\end{eqnarray}
The following expressions for the first and second order perturbations
of $\sqrt{-G}$ are also usefull
\begin{eqnarray}
\delta\sqrt{-G}_{(1)}&=&-\sqrt{-\bar{G}}\;
\bar{\Omega}_{jC}\;^{C}\Phi^{j}_{(1)}\\
\delta\sqrt{-G}_{(2)}&=&-\sqrt{-\bar{G}}\left(
\bar{\Omega}_{jC}\;^{C}\Phi^{j}_{(2)}+
(\bar{D}_{k}\;^{jC}\bar{D}_{jiC}\Phi^{i}_{(1)})
\Phi^{k}_{(1)}+\frac{1}{2}(\bar{D}^{k}\;_{jC}
\Phi^{j}_{(1)})(\bar{D}_{ki}\;^{C}\Phi^{i}_{(1)})\right. \nonumber\\
&-&\frac{1}{2}(\bar{\Omega}_{jC}\;^{C}\bar{\Omega}_{iD}\:^{D}-
\bar{\Omega}_{jCD}\bar{\Omega}_{i}\:^{CD})
\Phi^{j}_{(1)}\Phi^{i}_{(1)}\left. \right)
\end{eqnarray}

\end{document}